# Bifurcation of the earthquake source at the end of the Omori epoch

A. Guglielmi[1], O. Zotov[1,2]

[1] *Schmidt Institute of Physics of the Earth, Russian Academy of Sciences, Moscow, Russia, guglielmi@mail.ru, zavyalov@ifz.ru*

[2] *Borok Geophysical Observatory, the Branch of Schmidt Institute of Physics of the Earth, Russian Academy of Sciences, Borok, Yaroslavl Region, Russia, klb314@mail.ru, ozotov@inbox.ru*

**Abstract**

The earthquake source after the main shock can theoretically be represented as a black box without an entrance. At the output, there is a signal in the form of aftershocks, the frequency of which decreases on average with time according to the Omori law. The task of the researcher is to evaluate the structure, state and mechanism of functioning of a dynamic system simulating the earthquake source based on the output signal. In this paper, we outline an approach to a partial solution of this general problem. Omori's law is presented as a differential equation of aftershock evolution. An inverse problem has been posed and solved, the essence of which is to determine the source deactivation coefficient from the observed frequency of aftershocks. The existence of the so-called Omori epoch, during which the deactivation coefficient remains constant, has been discovered. At the end of the Omori epoch, the deactivation coefficient experiences complex variations. A hypothesis has been put forward that the end of the Omori epoch indicates a bifurcation, i.e. about the transition of the source from one state to a qualitatively different state. Our paper is written for the 100th anniversary of the death of Fusakichi Omori.

*Keywords*: aftershocks, Omori law, evolution equation, deactivation coefficient, inverse problem, Omori epoch, bifurcation point.

## 1. Introduction: Omori Epoch

In earthquake physics, the Omori law [1] is often mentioned, which states that the frequency of aftershocks $n$ decreases hyperbolically with time (see, for example, monograph [2] and reviews [3–5]). In differential form, Omori's law is

$$\frac{dn}{dt} + \sigma n^2 = 0. \qquad (1)$$

Here $\sigma$ is the source deactivation coefficient after the main shock of an earthquake [6]. Using (1), it is possible to pose the inverse problem of the earthquake source: to determine the deactivation coefficient from the observational data on the frequency of aftershocks. The solution of the inverse problem has the form:

$$\sigma = \frac{d}{dt}\langle g \rangle. \qquad (2)$$

Here $g = 1/n$ is the auxiliary function, the angle brackets denote regularization, which reduces to smoothing the auxiliary function.

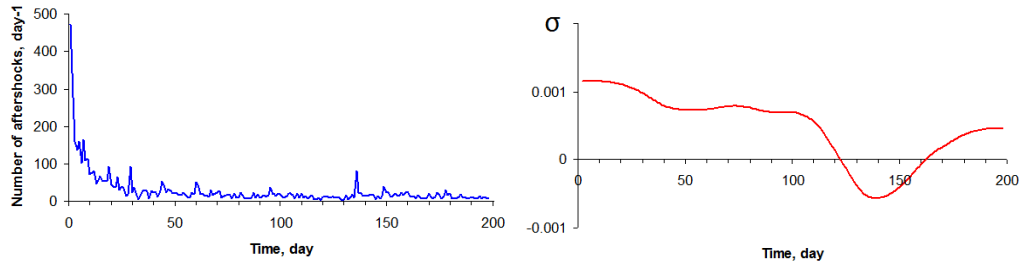

**Fig. 1**. Frequency of aftershocks (left) and source deactivation factor (right) after an earthquake with magnitude M = 5.4 at a depth of H=7 km in Northern California on 1983.01.07.

A typical pattern of aftershock frequency variation over time is shown in Figure 1 on the left. Shown is an event that occurred in Northern California on 1983.01.07.01h38m10s. The magnitude of the main shock M = 5.4, the depth of the hypocenter H=7 km according to the earthquake catalog http://www.ncedc.org. The result of solving the inverse problem is shown on the right. We see that the deactivation coefficient remains virtually unchanged for about 20 days after the mainshock. We call such a time interval the *Omori epoch*, since at $\sigma = $ const, the evolution of aftershocks strictly follows the hyperbolic Omori law [1].

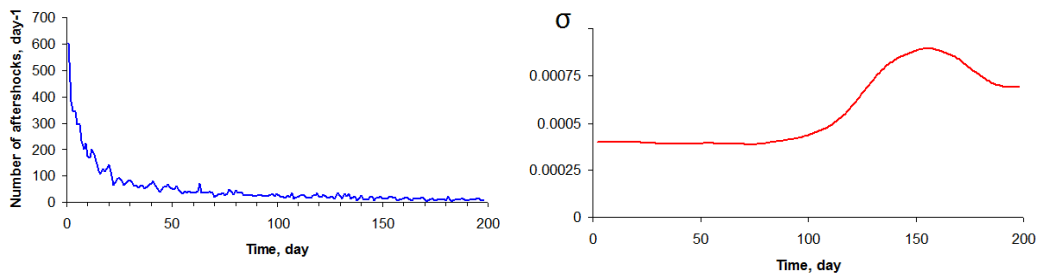

**Fig. 2**. Frequency of aftershocks (left) and source deactivation factor (right) after an earthquake with magnitude M = 6.7 at a depth of H=18 km in Southern California on 1994.01.17.

Our experience indicates that the duration of the Omori epoch varies from case to case approximately within 10–100 days [7, 8]. An example of a very long Omori epoch is shown in Figure 2. The event occurred in Southern California. The magnitude of the main shock M = 6.7, the depth of the hypocenter H=18 km according to the catalog https://scedc.caltech.edu.

In this paper, we want to discuss the idea that the end of the Omori epoch indicates the transition of the earthquake source as a dynamic system from one state to a qualitatively different state. The transition is fast in the sense that the transition period is much shorter than the duration of the Omori epoch. If we take into account that the continuous function $\sigma(t)$, which characterizes the source, is artificially smoothed in the process of regularization when solving the inverse problem, then the hypothesis of a sharp (jump-like) change in the state of the source as a dynamic system is quite plausible.

In Section 2, we present the result of an analysis of eight events, which clearly confirms the idea of a critical change in the state of the source after the end of the Omori epoch. In Section 3, we will discuss the question of interpretation of the phenomenon. Since we do not know the source equation of state, we will confine ourselves to a discussion of two possible ways, following which one can understand, at least in general terms, the rapid transition of the source from one state to another.

## 2. Bifurcation

The papers [7, 8] present the result of solving the inverse source problem for a small set of earthquakes. We use this set to support the idea that the end of the Omori epoch indicates a qualitative change in the state of the source. Eight events selected for analysis are presented in the Table. The pre-processing was that for each event we calculated the difference between $\sigma(t)$ and $\sigma(0)$. The resulting value $\triangle\sigma$ is plotted along the vertical axis in Figure 3. In addition, we performed synchronization by aligning the zero point in time in Figure 3 with the end of the Omori epoch.

**Table**. Information for Figure 3

| Region | Time, GMT | Magnitide | Lat, Lon, degrees | Depth, km | σ (Omori epoch) | Omori epoch, day |
|---|---|---|---|---|---|---|
| NC | 1983.01.07 01h38m10s | 5.4 | 37.64 -118.89 | 7 | 0.00114 | 23 |
| NC | 1983.05.02 23h42m38s | 5.4 | 36.23 -120.32 | 10 | 0.00167 | 18 |
| NC | 1984.04.24 21h15m18s | 6.4 | 37.31 -121.68 | 8 | 0.00104 | 28 |
| NC | 1984.11.23 18h08m25s | 6.0 | 37.46 -118.61 | 10 | 0.00061 | 63 |
| SC | 1986.07.08 09h20m44s | 5.6 | 33.99 -116.61 | 10 | 0.00174 | 21 |
| NC | 1989.10.18 00h04m15s | 7.0 | 37.04 -121.88 | 17 | 0.00096 | 75 |
| SC | 1994.01.17 12h30m55s | 6.7 | 34.21 -118.54 | 18 | 0.0004 | 100 |
| SC | 1999.10.16 09h46m44s | 7.1 | 34.59 -116.27 | 0.01 | 0.00024 | 98 |

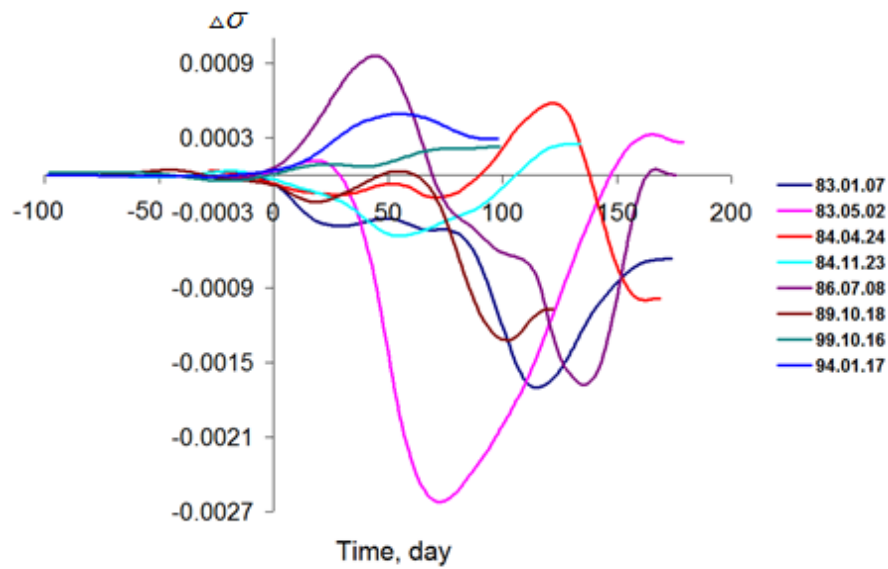

**Fig. 3**. Summary bifurcation diagram indicating a change in the deactivation mode of the earthquake source after the end of the Omori epoch (see text). The vertical axis shows the deactivation factor minus the value of this factor at the beginning of the Omori epoch. The critical point is aligned with zero point in time.

The result of normalization and synchronization, shown in Figure 3, strongly resembles the bifurcation phenomenon known in the theory of nonlinear dynamical systems. In the imaginary phase space of a dynamical system, the functioning of which presumably imitates the evolution of the source, the bifurcation point is reached at the end of the Omori epoch and is accompanied by a jump in the derivative $\theta = d\sigma/dt$ from values close to zero to positive or negative finite values. Our hypothesis is that there is a transition of the source from one state to a qualitatively different state at the critical point. The transition through a singular point can be performed both with an increase and a decrease in the deactivation coefficient.

### 3. Discussion

At this stage of the study, we still do not understand the mechanism of the rapid rearrangement of the source at the end of the Omori epoch. Perhaps, under the influence of an intense flow of aftershocks inelastic deformations gradually accumulate and a modification of the rock mass in the source suddenly occurs. However, this is nothing more than a general vague judgment, since the nature of the modification is not clear to us.

Let's try to formalize the problem. Equation (1) describes the source as a one-parametric one-dimensional system. We need either to introduce an additional (controlling) parameter, a small change of which leads to a qualitative metamorphosis of the source, or to increase the dimension of the system.

Let's consider the first possibility. Faraoni [9] drew attention to the fact that the Omori law (1) can be represented as the Euler-Lagrange equation. Guglielmi and Klain [10] (see also [5]) modified the Faraoni's Lagrangian and arrived at the Verhulst logistic equation

$$\frac{dn}{dt} = n(\gamma - \sigma n), \tag{3}$$

which contains the desired control parameter $\gamma$. In the Omori epoch, when $\sigma = c$onst and $n \gg n_\infty = \gamma/\sigma$, equation (3) is almost indistinguishable from (1). At the end of the aftershock process, seismicity goes into the background, $n = n_\infty$.

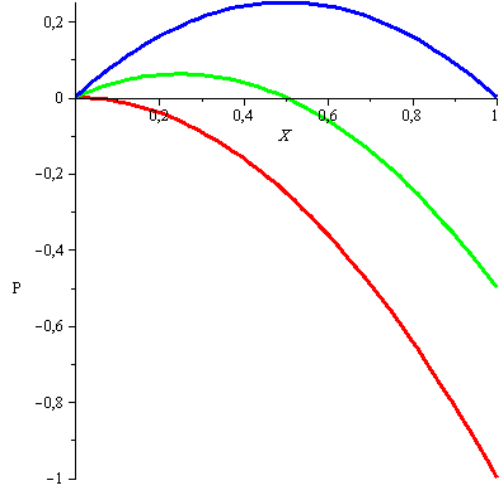

**Fig. 4**. Phase portrait on the phase plane of equation (3). The following notation is used here: $X = n/n_{max}$, $X_\infty = n_\infty/n_{max}$, $P = dX/dT$, $T = \sigma n_{max} t$. The red, green and blue phase trajectories are plotted at $X_\infty = 0$, 0.5 and 1, respectively (see text).

Faraoni proposed to introduce the phase plane $(n, \dot{n})$ to represent the trajectory of the dynamical system (1), imitating the evolution of the source according to the Omori hyperbolic law. In Figure 4, the Faraoni's phase portrait is shown by the red line. The portrait is the only parabola with a vertex at a stationary point (0, 0). It is easy to verify that this point is stable. The representing point moves in the phase plane from the initial state to the equilibrium state (0, 0) with a variable velocity in such a way that hyperbolic Omori law is strictly observed on the entire segment of the phase trajectory.

Above the red line in Figure 4 is one of the family of phase trajectories of the logistic equation (3). We find a radical change in the phase portrait. First, the stationary point (0, 0) becomes unstable. Secondly, a second stationary point (0, 0.5) appears, which is the point of stable equilibrium. We emphasize that the idealized trajectories of aftershocks (i.e., at $\sigma = c$onst, $\gamma = c$onst) correspond to segments of phase trajectories lying below the horizontal line both in the portrait of Faraoni ($\gamma = 0$) and in the portrait of Guglielmi-Klain ($\gamma > 0$).

Let us assume that the representative point starts moving along the Faraoni trajectory ($\gamma = 0$), i.e. along the red line in figure 4. However, at a certain moment the control parameter $\gamma$, for one reason or another, acquires a certain positive value. The state of a dynamical system will change radically, even if gamma is arbitrarily small. The representing point will make a transition from one phase trajectory to another, and the deactivation coefficient will decrease. On the contrary, with the reverse transition, the deactivation coefficient will increase. The presented picture gives some insight into the possible role of control parameter in the mechanism of the end of Omori epoch.

Thus, we have used the aftershock evolution equation (3) in order to shift the focus from the problem of the appearance of a bifurcation point to the problem of the stability of control parameter in the Omori epoch. This new problem is more accessible for analysis than the problem of the specific evolution of the deactivation coefficient after passing the critical point. It can, for example, be assumed that after the main shock and until the end of Omori period, the rocks in the source remain in a more or less consolidated state. As a result, the control parameter is stable (possibly equal to zero). And only under the action of a powerful flow of aftershocks, sooner or later, the geological environment in the source is modified, the control parameter undergoes a jump, which leads to the deviation of the aftershock frequency from the strictly hyperbolic Omori law.

We tried to understand the phenomenon of bifurcation by introducing a control parameter into the evolution equation (1). Now let's discuss another possibility. Namely, taking equation (1) as a basis, let us consider the question of whether it would not be reasonable to add to it a second equation describing the evolution of the deactivation coefficient. It would be possible to choose an appropriate equation in the theory of nonlinear dynamical systems, but we believe that the form of the equation should be prompted by experiment. Looking ahead, we say that in this work the desired equation will not be found. But we will indicate the path on which, in principle, it is possible to achieve the goal. The essence of our proposal is to apply the inverse problem method of the theory of nonlinear dynamic systems [11] to analyze the flow of aftershocks.

We will consider the earthquake source as a black box without an entrance. The structure and nature of the functioning of a dynamic system enclosed in a box must be estimated from only one output signal, namely, from the variation in the frequency of aftershocks. The question immediately arises about the dimension of the phase space of the system and about the choice of canonical variables. As a trial step, we choose the canonical variables $g$, $\sigma = dg/dt$ and consider the projection of the phase trajectory in the phase space of indeterminate dimension onto the plane ($g$, $\sigma$).

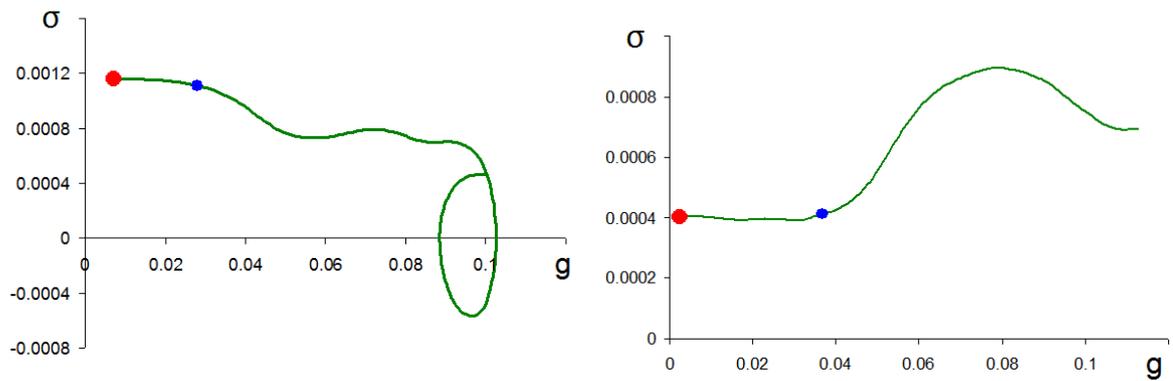

**Fig. 5**. Phase trajectories in the plane ( $g$ , $\sigma$ ). The left (right) shows the trajectory for the event depicted in Figure 1 (Figure 2). Red dots indicate the start of the path. Blue dots indicate the end of the Omori epoch.

Figure 5 on the left (right) shows the result of the analysis for the event depicted in Figure 1 (Figure 2). We again see that the end of the Omori epoch is accompanied by a qualitative change in the functioning of the source, but this is perhaps all that can be extracted from the analysis in the ( $g$ , $\sigma$ ) plane.

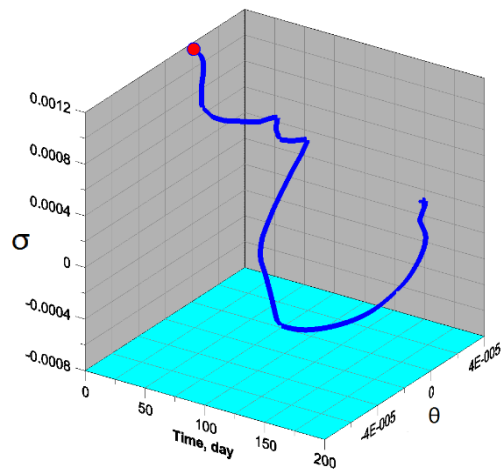

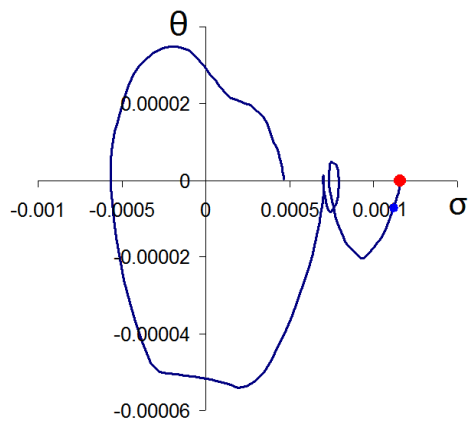

**Fig. 6**. The upper figure shows the movement of the representative point in the coordinates $\sigma(t)$, $\theta(t)$ for the event shown in Figure 1. The lower figure shows the corresponding trajectory on the phase plane ($\sigma, \theta$). The red dot marks the beginning of the movement. The blue dot marks the end of the Omori epoch on the 20th day of the evolution of the dynamical system.

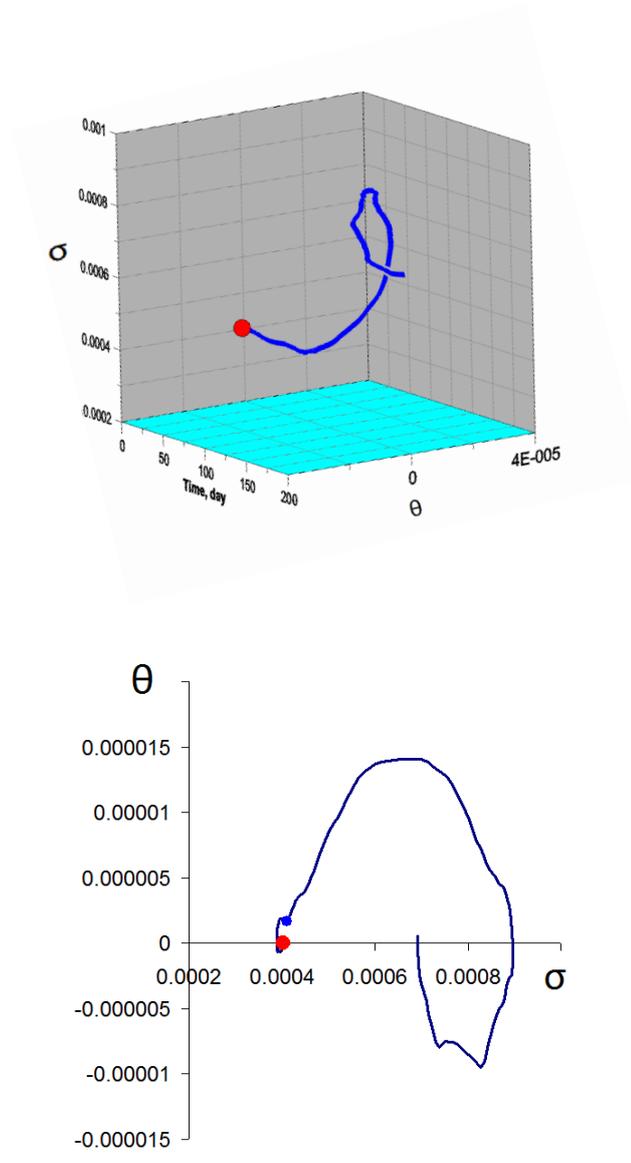

**Fig. 7**. The upper figure shows the movement of the representative point in the coordinates $\sigma(t)$, $\theta(t)$ for the event shown in Figure 2. The lower figure shows the corresponding trajectory on the phase plane ($\sigma, \theta$). The red dot marks the beginning of the movement. The blue dot marks the end of the Omori epoch on the 90th day of the evolution of the dynamical system.

We got a more meaningful picture when choosing the canonical variables $\sigma$, and $\theta = d\sigma/dt$. Figures 6 and 7 show the result of analyzing the same two events shown in Figure 5. Three-dimensional trajectories are shown here, with the time $t$ being considered as a parameter. The projections of the trajectories onto the plane ($\sigma$, $\theta$) are also shown. On the 3D graphs, we see that after the end of Omori epoch, the trajectories have curvature and torsion. Two-dimensional phase portraits are also quite expressive. We propose to continue this kind of analysis in order, having accumulated experience, to try to guess the form of equation describing the evolution of deactivation coefficient. The richness and diversity of the phase portraits, three of which are shown in Figure 8, inspire hope for success in the implementation of our plan.

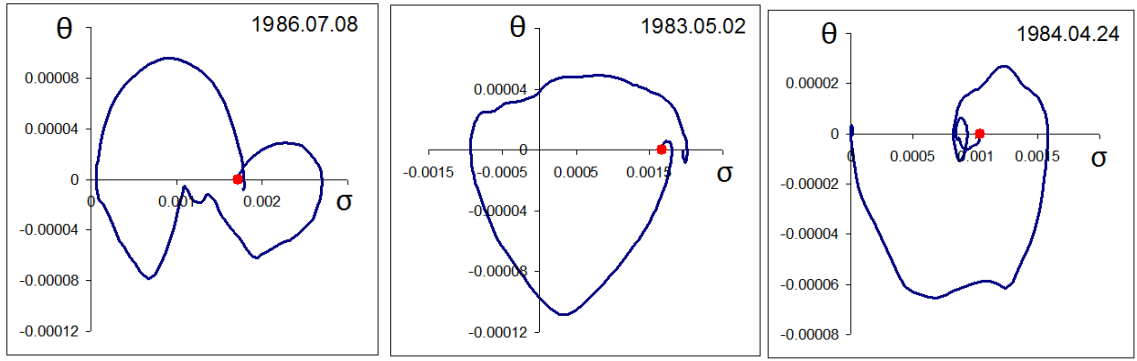

**Fig. 8.** Examples of phase portraits of the earthquake source.

Concluding the discussion of the bifurcation problem, we would like to draw attention to the fact that the problem, generally speaking, fits into the program for constructing the geometrodynamics of the earthquake source outlined in [12]. In this work, a list of problems of source physics is given, in which geometric images and constructions play a key role in the representation of geodynamic processes that accompany a strong earthquake. Phase trajectories, two-dimensional and three-dimensional phase portraits naturally fit into the context of geometrodynamics.

## 4. Conclusion

We consider the discovery of a bifurcation point in the dynamics of aftershocks as an interesting new result of a long-term study of the earthquake source. It significantly complements other interesting results obtained earlier by the inverse problem method based on the evolution equation (1): the existence of the Omori epoch, the perturbation of the hyperbolic Omori law by the round-the-world seismic echo and free oscillations of the Earth, the concept of the source proper time, the existence of mirror triads, and others (see, for example, [4, 5, 13, 14]).

It seems appropriate for us to briefly recall here some stages in the development of aftershocks theory from the discovery of Omori's law in 1894 to the present day. In 1924, Hirano suggested replacing Omori's two-parameter hyperbolic formula $n(t) = k/(c+t)$ [1] with a three-parameter power function $n(t) = k/(c+t)^p$ [15]. In the second half of the last century, Utsu introduced Hirano's methodological approach into wide use (see, for example, review [3] and the literature cited therein). In many works, the exponent $p$ was calculated from the observational data on the frequency of aftershocks $n(t)$ after earthquakes that occurred in various geological conditions. Sometimes, not only Hirano's law, but also Omori's law was considered as simply fitting formulas, allowing a compact representation of observational data.

In contrast, our research group [5, 12, 14] considered the Omori law [1] as a fundamental law of nature. The discovery of a rather long Omori epoch can be regarded as confirmation of the correctness of our methodological approach.

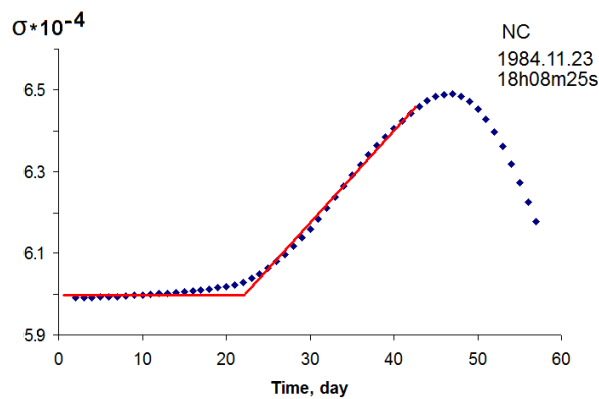

Fig. 9. Schematic representation of the search for a bifurcation point. The dotted curve shows the variation in the deactivation factor. The solid broken line shows the fitting function.

The discovery of the bifurcation phenomenon at the end of the Omori epoch testifies to the effectiveness of our one-parameter formulation of the Omori law (1). It is quite easy to determine the bifurcation point, as shown schematically in Figure 9. It is also easy to determine the time derivative of the fitting function. The plan of our further work includes the task of finding the dependence of the jump in the derivative on the magnitude of the mainshock, the value of the deactivation coefficient in the Omori epoch, and on other source parameters.

In conclusion, we would like to remind that this year marks the 100th anniversary of the death of Fusakichi Omori [16]. He is honored to be the author of the discovery of the law of aftershock evolution. With the discovery of Omori's law, made 130 years ago, the modern stage of development of earthquake physics actually began.

*Acknowledgments*. This work was done in close creative collaboration with B.I. Klain and A.D. Zavyalov. We express our deep gratitude to them. The work was carried out according to the plan of state assignments of IPhE RAS. We are grateful to the compilers of the earthquakes catalogs of Southern (https://scedc.caltech.edu) and Northern (http://www.ncedc.org) California, data from which were used in our study.